\documentclass[aps,prb,floatfix,twocolumn,showpacs,preprintnumbers,amsmath,amssymb,superscriptaddress]{revtex4-2}
\usepackage{graphicx}
\usepackage{bm}
\usepackage{braket}
\usepackage{epstopdf}

\begin{document}
\title{Anisotropic topological superconductivity in Josephson junctions}
\author{Bar{\i}\c{s} Pekerten}
\author{Joseph D. Pakizer}
\author{Benjamin Hawn}
\author{Alex Matos-Abiague}
\affiliation{Department of Physics and Astronomy, Wayne State University, Detroit, MI 48201, USA}

\date{\today}
 
\begin{abstract}
We investigate the effects of magnetic and crystalline anisotropies on 
the topological superconducting state of planar Josephson junctions 
(JJs). In junctions where only Rashba spin-orbit coupling (SOC) is 
present, the topological phase diagram is insensitive to the 
supercurrent direction, but exhibits a strong dependence on the 
magnetic field orientation. However, when both Rashba and Dresselhaus 
SOCs coexist, the topological phase diagram strongly depends on both 
the magnetic field and junction crystallographic orientations. We 
examine the impact of the magnetic and crystalline anisotropy on the 
current-phase relation (CPR), energy spectrum, and topological gap of 
phase-biased JJs, where the junction is connected in a loop and the 
superconducting phase difference is fixed by a loop-threading magnetic 
flux. The anisotropic CPR can be used to extract the ground-sate phase 
(i.e. the superconducting phase difference that minimizes the system 
free energy) behavior in phase-unbiased JJs with no magnetic flux. 
Under appropriate conditions, phase-unbiased JJs can self-tune into or 
out of the topological superconducting state by rotating the in-plane 
magnetic field. The magnetic field orientations at which topological 
transitions occur strongly depend on both the junction 
crystallographic orientation and the relative strength between Rashba 
and Dresselhaus SOCs. We find that for an optimal practical 
application, in which the junction exhibits topological 
superconductivity with a sizable topological gap, a careful balancing 
of the magnetic field direction, the junction crystallographic 
orientation, and the relative strengths of the Rashba and Dresselhaus 
SOCs is required. We discuss the considerations that must be undertaken 
to achieve this balancing for various junction types and parameters.
\end{abstract}

%\pacs{74.78.Na,74.25.Ha,74.45.+c}
\maketitle

\vspace{-.2cm}
\section{Introduction}
\vspace{-.2cm}

Majorana bound states (MBSs) are zero-energy quasiparticle excitations 
predicted to arise in topological superconductors 
(TSs).~\cite{Kitaev2001:PU, Kitaev2003:AP, Leijnse2012:SST, 
Beenakker2013:ARCMP, Aguado2017:RNC, Qi2011:RMP} Due to their 
non-Abelian exchange statistics they can be utilized as qubits for 
fault-tolerant quantum computing, with quantum gates realized through 
braiding operations.~\cite{Ivanov2001:PRL,Nayak2008:RMP, Alicea2011:NP, 
Aasen2016:PRX} Driven by this technological impetus, proposals to 
achieve topological superconductivity have included, among others, 1D 
systems such as magnetic chains on \textit{s}-wave 
superconductors,~\cite{Choy2011:PRB, Martin2012:PRB, Pientka2013:PRB, 
NadjPerge2013:PRB, NadjPerge2014:S,Pawlak2016:NPJQI} semiconductor 
nanowires with large spin-orbit coupling (SOC) proximitized by 
\textit{s}-wave superconductors,~\cite{Oreg2010:PRL, Sau2010:PRB, 
Sau2010:PRL, Lutchyn2010:PRL, Rokhinson2012:NP, Pientka2012:PRL, 
Mourik2012:S, Das2012:NP,Deng2012:NL, Deng2016:S, Manna2020:PNAS} and 
proximitized systems exposed to magnetic textures.~\cite{Fatin2016:PRL, 
MatosAbiague2017:SSC, Mohanta2019:PRA, Zhou2019:PRB, Klinovaja2012:PRL, 
Kjaergaard2012:PRB, Marra2017:PRB, Desjardins2019:NM, 
Steffensen2021:PRB} Because of their experimental feasibility, planar 
Josephson junctions (JJs) have also been considered as an alternative 
promising platform for creating and manipulating 
MBSs.~\cite{Pientka2017:PRX, Setiawan2019:PRB, Setiawan2019:PRB2, 
Fornieri2019:N,Ren2019:N,Dartiailh2021:PRL, Hart2014:NP, 
Laeven2020:PRL, Lesser2021:PNAS, Zhou2021:arxiv, Zhou2020:PRL, 
Scharf2019:PRB, Cayao2017:PRB, Kontos2002:PRL, Yokoyama2014:PRB, 
Hell2017:PRL, Zhang2020:PRB, Woods2020:PRB, Stenger2019:PRB, 
Svetogorov2021:PRB, Pekker2013:PRL, Paudel2021:PRB, Salimian2021:APL, 
Setiawan2017:PRB, Setiawan2021:arXiv, Banerjee2022:arXiv}  Moreover, 
the superconducting phase difference across JJs provides an additional 
control knob that can enhance the parameter space leading to 
topological superconductivity. However, tuning a planar JJ to a 
topologically nontrivial state does not necessarily guarantee the 
existence of a sizable topological gap, which is a practical 
requirement for the stability of MBSs and braiding 
operations.~\cite{Tewari2010:AoP, Alicea2010:PRB} Indeed, as shown 
later in this work (see also Ref.~\onlinecite{Pakizer2021:PRR}), even 
when the parameter space for topological superconductivity is 
relatively large, the topological gap may be sizable only over reduced 
subregions.

Josephson junctions with noncentrosymmetric superconductors 
(particularly $d$-wave superconductors) have been predicted to exhibit 
anisotropic effects.~\cite{Tanaka1996:PRB, Tanaka1997:PRB, 
Tanaka2010:PRL} In this work, we consider the effects of magnetic and 
crystalline anisotropies on the topological superconducting state in 
planar JJs formed in a semiconducting two-dimensional electron gas 
proximitized by $s$-wave superconductors. The interrelation between the 
Zeeman interaction and the Rashba SOC emerging from the lack of 
structure inversion symmetry~\cite{Bychkov1984:JPC} in proximitized 
planar JJs gives rise to a strong dependence of the system properties 
on the magnetic field direction. Furthermore, in junctions where both 
Rashba and Dresselhaus SOCs are relevant, not only the magnetic field 
direction, but also the junction crystallographic orientation can 
strongly affect the topological superconducting 
state.~\cite{Pakizer2021:PRR} The Dresselhaus SOC originates from the 
bulk inversion asymmetry,~\cite{Dresselhaus1955:PR} which can be 
particularly large in some zinc blende semiconductors (e.g., InSb) 
suitable for building superconductor/semiconductor proximitized 
JJs.~\cite{Mayer2020:AEM} Here we investigate the impact of SOC-induced 
anisotropies of topological phase transitions on the topological gap, 
topological charge, energy spectrum, ground-state phase, current phase 
relation, and critical currents in planar JJs.

\begin{figure}[t]
\centering
\includegraphics*[width=0.95\columnwidth]{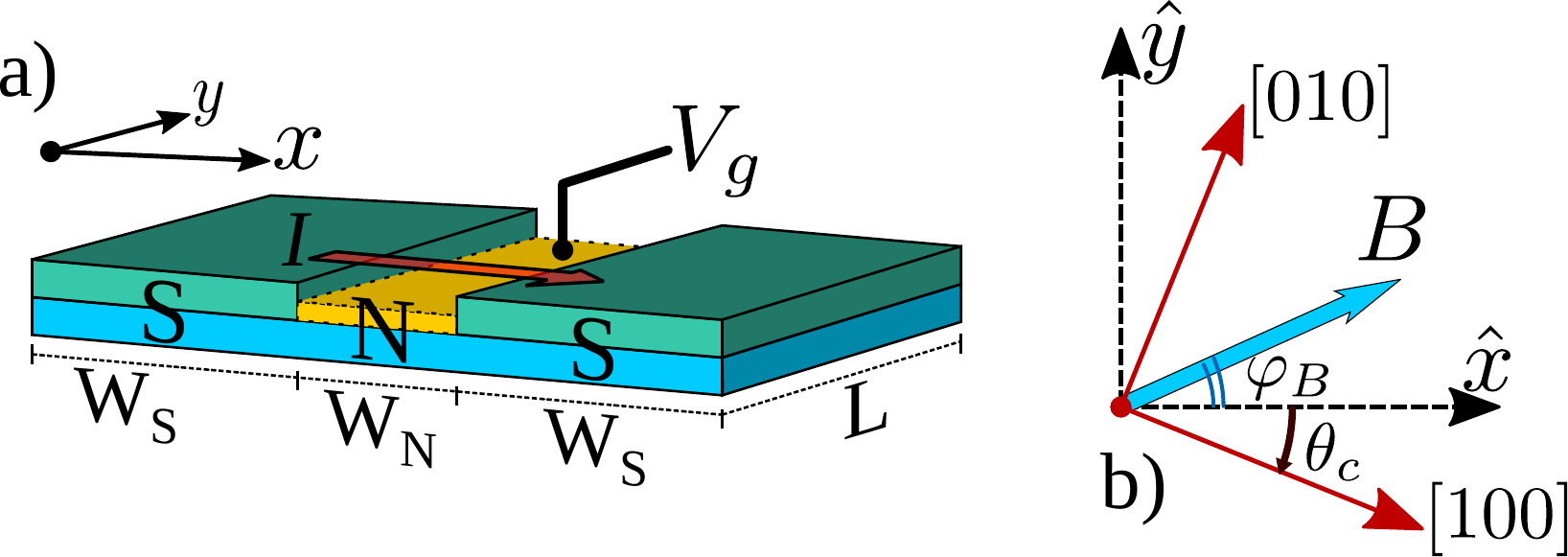}
\caption{[Color online] (a) A JJ composed of a non-centrosymmetric 
semiconductor 2DEG in contact to two superconducting (S) leads. The 
$\hat{x}$ and $\hat{y}$ axes define the coordinate system in the 
junction reference frame. The Rashba SOC strength can be controlled by 
using a gate on the top of the normal region~\cite{Dartiailh2021:PRL, 
Mayer2020:NC}. The current flow is perpendicular to the junction. (b) 
$\varphi_B$ and $\theta_c$ characterize the orientation of the in-plane 
magnetic field ($\mathbf{B}$) and the junction reference frame, 
respectively, with respect to the semiconductor [100] crystallographic 
axis. }
\label{fig:jj}
\end{figure}

\section{Theoretical Model}

\subsection{General considerations}

We consider JJs composed of superconducting (S) and normal (N) regions 
(see Fig.~\ref{fig:jj}). The S regions are formed in a semiconducting 
2DEG proximitized by a superconducting (e.g., Al or Nb) covering. 
Excitations in the JJ are described by the Bogoliubov-de Gennes (BdG) 
Hamiltonian,
\begin{equation}\label{H-BdG}
H=H_{0}\tau_z-\frac{g^\ast\mu_B}{2}\mathbf{B}\cdot\boldsymbol{\Sigma}+\Delta(x)\tau_+ +\Delta^\ast(x)\tau_-\;,
\end{equation}
where
\begin{eqnarray}\label{Ho}
H_{0}&=&\frac{\mathbf{p}^2}{2m^\ast}+V(x)-(\mu_S-\varepsilon)+\frac{\alpha}{\hbar}\left(p_y\sigma_x - p_x\sigma_y\right)\\ \nonumber
&+&\frac{\beta}{\hbar}[(p_x\sigma_x - p_y\sigma_y)\cos2\theta_c - (p_x\sigma_y + p_y\sigma_x)\sin2\theta_c]\;,
\end{eqnarray}
is the single-particle Hamiltonian of the 2DEG in the absence of a 
magnetic field. In the equations above, $\mathbf{p}$ is the momentum, 
$m^\ast$ the electron effective mass, $\alpha$ and $\beta$ are, 
respectively, the Rashba and Dresselhaus SOC strengths, and $\theta_c$ 
characterizes the direction of the current ($x$ axis) with respect to 
the [100] crystallographic direction of the semiconductor 
[Fig.~\ref{fig:jj}(b)]. The crystallographic orientation of the 
junction is determined by the angle $\theta_c+\pi/2$. The length of the 
junction is $L$ and the widths of the S and N regions are $W_S$ and 
$W_N$, respectively (see Fig.~\ref{fig:jj}). The gate-voltage-induced 
difference between the chemical potentials in the N ($\mu_N$) and S 
($\mu_S$) regions is described by $V(x)=(\mu_S - 
\mu_{N})\,\Theta(W_N/2-|x|)$ and $\sigma_{x,y,z}$ and $\tau_{x,y,z}$, 
with $\tau_\pm =(\tau_x\pm i\tau_y)/2$, represent Pauli and Nambu 
matrices, respectively. The chemical potentials are measured with 
respect to the minimum of the single-particle energies, 
$\varepsilon=m^\ast\lambda^2(1+\left|\sin 2\theta_c\right|)/2\hbar^2$,  
where we have used the SOC parametrization,
\begin{equation}\label{param}
\alpha=\lambda\cos\theta_{so}\;,\;\;\beta=\lambda\sin\theta_{so}\;,\;\;\lambda=\sqrt{\alpha^2+\beta^2}.
\end{equation}
Here $\lambda$ represents the overall strength of the combined Rashba + 
Dresselhaus SOCs, while the spin-orbit angle,
\begin{equation}\label{that-so}
\theta_{so}={\rm arccot}(\alpha/\beta),
\end{equation}
characterizes the relative strength between Rashba and Dresselhaus SOC.

The second contribution in Eq.~(\ref{H-BdG}), with the Dirac spin 
matrices $\boldsymbol{\Sigma}$, represents the Zeeman splitting due to 
an applied magnetic field,
\begin{equation}\label{def-B}
\mathbf{B}=|\mathbf{B}|
\begin{pmatrix}
\cos\varphi_B \\
\sin\varphi_B \\
0
\end{pmatrix}.
\end{equation}
The angle $\varphi_B$ characterizes the direction of the magnetic field 
with respect to the current flow ($x$ axis), as shown in 
Fig.~\ref{fig:jj}(b). The spatial dependence of the superconducting gap 
is $\Delta(x)=\Delta e^{-i\;{\rm sgn}(x)\phi/2}\,\Theta(|x|-W_N/2)$, 
where $\phi$ is the phase difference across the JJ. 

The temperature and magnetic field dependence of the superconducting 
gap is taken into account by using the BCS relation,
\begin{equation}
\Delta(T,B)\approx \Delta(T,0)\sqrt{1-\left[\frac{B}{B_c(T)}\right]^2}, 
\end{equation}
where $\Delta(T,0)\approx \Delta_0\tanh[1.74\sqrt{T_c/T-1}]$, 
$\Delta_0 = 1.74$~$k_B T_c$, $k_B$ is the Boltzmann constant, and $T_c$ 
is the superconductor critical temperature. The temperature dependence 
of the critical magnetic field can be approximated as, 
$B_c(T)=B_c(1-T^2/T_c^2)$, where $B_c$ is the critical magnetic field 
at zero temperature.

Although we consider ballistic junctions throughout this work, we 
expect the predicted anisotropic effects to qualitatively hold in the 
presence of weak disorder. Junctions with low disorder are usually 
preferred because the presence of disorder typically reduces the 
parameter space supporting topological superconductivity, although in 
some cases weak disorder can increase the robustness of the topological 
superconducting state~\cite{Rieder2013:PRB, Adagideli2014:PRB, 
Pekerten2017:PRB, Haim2019:PRL}.

\subsection{Topological gap and topological charge}

Topological superconductivity (TS) is a superconducting phase featuring 
a pair of degenerate zero-energy states, called Majorana bound states 
(MBSs), which are isolated from the rest of the excitation spectrum by 
an energy gap, called the topological gap ($\Delta_{top}$). In the TS 
state, the topological gap cannot be destroyed by smooth local 
perturbations, providing protection for the MBSs. However, the 
information stored in the MBSs can be damaged if the perturbation 
energy becomes comparable or larger than the topological gap. A large 
topological gap is therefore desirable for the practical use of 
fault-tolerant qubits encoded in MBSs.

The topological gap can be estimated by imposing translational 
invariance along the junction direction (the $y$ direction in our 
case). In such a system, the momentum component $p_y$ is a conserved 
quantity and can be substituted by $\hbar k_y$. Then $\Delta_{\rm top}$ 
is obtained as the eigenenergy closest to zero, % 
\begin{equation}\label{topogap} \Delta_{\rm top}=\min_{k_y}|E(k_y)|. 
\end{equation} % Note that this quantity represents the topological gap 
only when the system is in the TS state. When the system is in the 
trivial state, the quantity defined in Eq.~(\ref{topogap}) simply 
denotes the lowest positive-energy Andreev state.

The size of the topological gap strongly depends on the interrelation 
between the spin-orbit angle ($\theta_{so}$), the junction 
crystallographic orientation ($\theta_c$) and the in-plane magnetic 
field orientation ($\varphi_B$).~\cite{Scharf2019:PRB,Pakizer2021:PRR} 
In symmetric JJs (i.e., with identical left and right S coverings) the 
optimal topological gap is achieved when the system is in the TS state 
and the condition,~\cite{Pakizer2021:PRR}
\begin{equation}\label{opt-mag}
\tan{\varphi_B} = \cot\theta_{so}\; \sec{2\theta_c} - \tan{2\theta_c},
\end{equation}
is fulfilled. 

This work focuses on the behavior of the current across the JJ (see 
Fig.~\ref{fig:jj}a). Hence, we use the current direction as the axis 
with respect to which the magnetic field orientation, $\varphi_B$, is 
defined (see Fig.~\ref{fig:jj}b)~\cite{note1}.

In the presence of Rashba and Dresselhauss SOC the topological gap 
exhibits strong magnetic and crystalline 
anisotropies.~\cite{Pakizer2021:PRR} Therefore the fulfillment of 
Eq.~(\ref{opt-mag}) is a vital prerequisite for the optimization of 
$\Delta_{top}$. However, the relation in Eq.~(\ref{opt-mag}) alone is 
not sufficient for inducing TS. Therefore, in a practical situation, 
one will need to first arrange the experimental setup in a way that 
Eq.~(\ref{opt-mag}) is fulfilled, and then tune other system parameters 
(e.g., chemical potential, magnetic field amplitude) to drive the 
system into the TS state.

The TS state in symmetric JJs typically belongs to the D class. 
However, the BDI class can emerge for some specific junction 
crystallographic orientations and magnetic field directions (see Table 
I in Ref.~\onlinecite{Pakizer2021:PRR}). Therefore, the transition 
between the trivial and TS states typically occur when the $Z_2$ 
topological index (also called topological charge, $Q$) associated with 
the symmetry class D changes sign. According to the bulk-boundary 
correspondence, we can obtain the phase diagram of a junction with 
finite length by computing the topological charge of the 
translational-invariant version of the junction, 
\begin{equation}\label{qtopo}
Q = {\rm sgn}\left[\frac{{\rm Pf}\{H(k_y = \pi)\sigma_y\tau_y\}}{{\rm Pf}\{H(k_y = 0)\sigma_y\tau_y\}}\right],
\end{equation}
where ${\rm Pf}\{...\}$ denotes the 
Pfaffian~\cite{Tewari2012:PRL,Schnyder2008:PRB, Ryu2010:NJP, 
Ghosh2010:PRB}. The topological charge determines whether the system is 
in the trivial ($Q=1$) or topological ($Q=-1$) 
phase.~\cite{Stanescu2011:PRB, Stanescu2013:JoPCM, Rieder2013:PRB, 
Adagideli2014:PRB, Pekerten2017:PRB, Pekerten2019:PRB}

\subsection{Current-phase relation and critical current in Josephson junctions}

The supercurrent across the JJ can be obtained from the energy spectrum 
of the BdG Hamiltonian given in Eq.~(\ref{H-BdG}). Indeed, the 
eigenenergies ($E_n$) can be used to compute the free energy of the 
junction,
\begin{equation}\label{EQN:free-energy}
F=-\sum_{E_n> 0}\left[E_n + 2 k_B T \ln\left(1+e^{-\beta E_n}\right)\right]
\end{equation}
with $\beta = 1/k_B T$. The current-phase relation (CPR) is then 
obtained as
\begin{equation}\label{EQN:current-def}
I(\phi) = \frac{2e}{\hbar}\frac{dF}{d\phi}=I_0(\phi)+\Delta I(\phi,T).
\end{equation}
where the zero-temperature contribution is given by
\begin{equation}\label{EQN:izero}
I_0(\phi) = -\frac{2e}{\hbar}\sum_{E_n> 0}\frac{dE_n}{d\phi},
\end{equation}
while the temperature-dependent correction reads,
\begin{equation}\label{EQN:ifinite}
\Delta I(\phi,T) =\frac{4e}{\hbar}\sum_{E_n> 0}\left[\frac{1}{1+e^{\beta E_n}}\right]\frac{dE_n}{d\phi}.
\end{equation}
In the low-temperature limit ($\beta E_n\gg 1$), 
Eqs.~(\ref{EQN:current-def})--(\ref{EQN:ifinite}) yield
\begin{eqnarray}\label{EQN:low-t}
	I(\phi)&\approx& I_0(\phi)+\frac{4e}{\hbar}\sum_{E_n> 0} e^{-\beta E_n}\frac{dE_n}{d\phi}\nonumber\\
	&=&I_0(\phi)-\frac{4e}{\hbar\beta}\frac{d}{d\phi}\left(\sum_{E_n>0}\ e^{-\beta E_n}\right),
\end{eqnarray}
while in the high-temperature regime ($\beta E_n\ll 1$),
\begin{equation}\label{EQN:high-t}
	I(\phi)\approx -\frac{e\beta}{\hbar}\sum_{E_n> 0}E_n\frac{dE_n}{d\phi}=-\frac{e\beta}{2\hbar}\frac{d}{d\phi}\left(\sum_{E_n>0}E_n^2\right).
\end{equation}

In the phase-biased case, the JJ is connected to a closed loop threaded 
by a magnetic flux, $\Phi$, and the superconducting phase difference 
across the junction is fixed to the value $\phi=2\pi\Phi/\Phi_0$, where 
$\Phi_0$ is the magnetic flux quantum. In this case the current-phase 
relation [Eq.~\ref{EQN:current-def}] can be experimentally measured by 
tuning the magnetic flux.

In the absence of a magnetic flux the junction is phase unbiased and 
the phase difference self-adjusts in such a way that the free energy of 
the system is minimized. The ground-state phase ($\phi_{GS}$) is the 
superconducting phase difference that minimizes the free energy of the 
system, i.e.,
\begin{equation}
    F(\phi_{GS})=\min_{\phi}F(\phi)
\end{equation}
and the ground-state spectrum is the energy spectrum of the JJ 
evaluated at the ground-state phase, i.e., $E_n(\phi_{GS})$. The 
mathematical conditions for the free energy to have a minimum at the 
ground-state phase are,
\begin{equation}
\left.\frac{dF}{d\phi}\right|_{\phi=\phi_{GS}}=0\;\;{\rm and}\;\;\left.\frac{d^2F}{d\phi^2}\right|_{\phi=\phi_{GS}}>0,
\end{equation}
which, according to Eq.~(\ref{EQN:current-def}) can be rewritten as,
\begin{equation}\label{i-phi-gs}
    I(\phi_{GS})=0\;\;{\rm and}\;\; \left.\frac{dI}{d\phi}\right|_{\phi=\phi_{GS}}>0.
\end{equation}
The relations above allow for extracting the ground-state phase from 
the CPR.

Under forward bias voltage, the critical current of a phase-unbiased JJ 
is obtained by maximizing the current amplitude with respect to the 
phase difference, i.e.,
\begin{equation}\label{EQN:ic-def-jj}
    I_{c}=\max_{\phi} I(\phi).
\end{equation}
However, under reverse bias voltage, the critical current is negative 
and is determined by minimizing the supercurrent. In centrosymmetric 
JJs the amplitudes of the forward and reverse critical currents are 
equal. However, the interrelation between the SOC and the in-plane 
magnetic field can break the inversion symmetry and lead to the 
so-called superconducting diode effect~\cite{Wakatsuki2017:SciAdv, 
Qin2017:NatCommun, Hoshino2018:PRB, Yasuda2019:NatCommun, Ando2020:N, 
Baumard2020:PRB, Baumgartner2022:NN}, where the amplitudes of the 
forward and reverse critical currents become different. In this work we 
limit our analysis to the case of the forward critical current.

\subsection{Numerical Approach}

We use a finite-difference discretization of Eq.~(\ref{H-BdG}) to build 
a tight-binding version of the BdG Hamitonian, which is then 
numerically diagonalized to find the eigenstates and energy spectrum 
(see Appendix~\ref{SECT:APP:TB_simulations} for more details). The 
Pfaffians of the tight-binding Hamiltonian with imposed translational 
invariance along the junction direction are numerically calculated to 
compute the topological charge, while the energy spectrum is used to 
calculate the free energy of the system, CPR, ground state phase, and 
critical currents. 

The numerical simulations of the tight-binding version of the BdG 
Hamiltonian are performed by using the Kwant 
package~\cite{Groth2014:NJP}. To illustrate and compare the different 
effects of magnetic and crystalline anisotropies, two type of junctions 
are considered: i) Al/HgTe JJs, where only Rashba SOC plays a role and 
ii) Al/InSb JJs, where both Rashba and Dresselhaus SOC become relevant 
(see Appendix for more details).

\section{Current-phase relation in phase-biased Josephson junctions}\label{SECT:PhaseBiasedJJ}

\subsection{Effects of magnetoanisotropy}

To investigate the effects of magnetic field orientation on the CPR of 
a phase-biased JJ, we consider Al/HgTe JJs (see system parameters in 
Appendix~\ref{SECT:APP:TB_simulations}), where Rashba SOC is large and 
Dresselhaus SOC is negligibly small. In such systems the spin-orbit 
angle $\theta_{so}=0$ and the CPR is independent of the junction 
crystallographic orientation. For the chosen system parameters, the 
estimated zero-field superconducting coherence length of the Al/HgTe JJ 
is $\xi=81$nm, which is smaller but comparable to the width of the 
normal region, and about 1/4th the size of each lead.

\begin{figure}[t]
\centerline{\includegraphics[width=1.05\columnwidth]{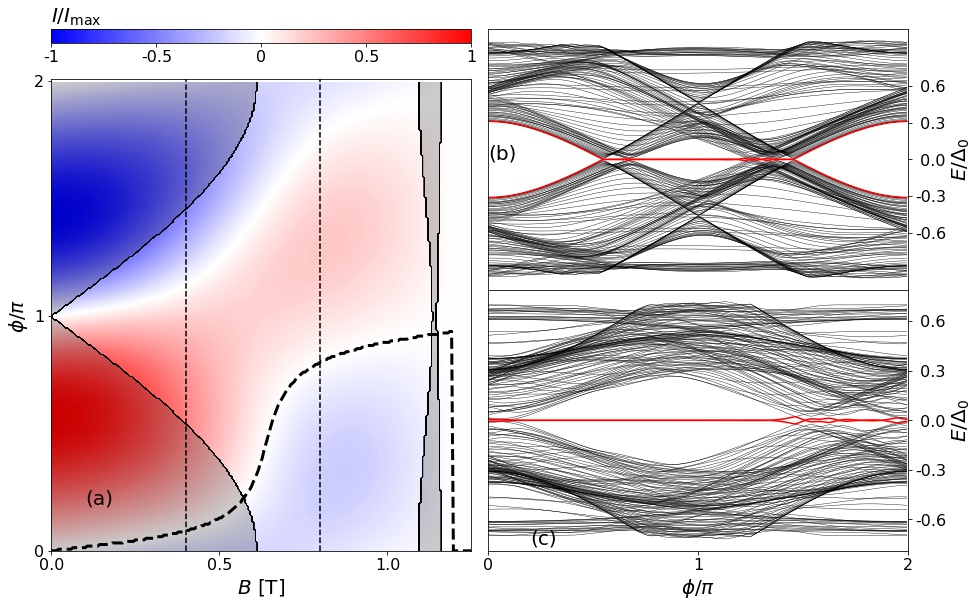}}
\caption{[Color online] (a) Plot of $I(\phi)$ and $Q$ as a function of 
$\phi$ and $B$ for a HgTe Josephson junction (see Figure~\ref{fig:jj}). 
Here, $\theta_\textrm{SO}=\theta_c=0$ and $\varphi_B=\pi/2$. The TB 
simulations are made for a translationally invariant 
system in the $y$-direction (Appendix~\ref{SECT:APP:TB_simulations}), 
with a SC lead width of $252$nm each and a junction width of $96$nm. 
The vertical dashed lines correspond to $B=0.4$T and $B=0.8$T (see 
Figure~\ref{fig:I_Q_vs_phi_vs_varphiB_BAniso}). The shaded (unshaded) 
areas have $Q=1$ ($Q=-1$). The TB lattice parameter is $a=6$nm, 
yielding a hopping parameter of $t=27.9$meV. (b) and (c) The lowest 200 
energy levels corresponding to a fully 2D TB simulation of the system 
in (a), with a length $L=4002$nm, at (b) $B=0.4$T and (c) $B=0.8$T. The 
red lines correspond lowest energy levels. }
\label{fig:I_Q_vs_phi_vs_Bz}
\end{figure}

The supercurrent (normalized to its maximum value) of a Al/HgTe JJ is 
shown in Fig.~\ref{fig:I_Q_vs_phi_vs_Bz}(a) as a function of an 
in-plane magnetic field perpendicular to the current (i.e., 
$\varphi_B=\pi/2$) and the superconducting phase difference. The shaded 
(unshaded) areas correspond to topological charge $Q=1$ (trivial state) 
and $Q=-1$ (TS state), respectively. Both the supercurrent amplitude 
and direction can be tuned by changing the phase difference and/or the 
magnetic field strength. Note that for magnetic fields larger than the 
critical field (1.19~T at 0.7~K), superconductivity is destroyed and 
the supercurrent vanishes.

\begin{figure}[t]
\centerline{\includegraphics[width=1.05\columnwidth]{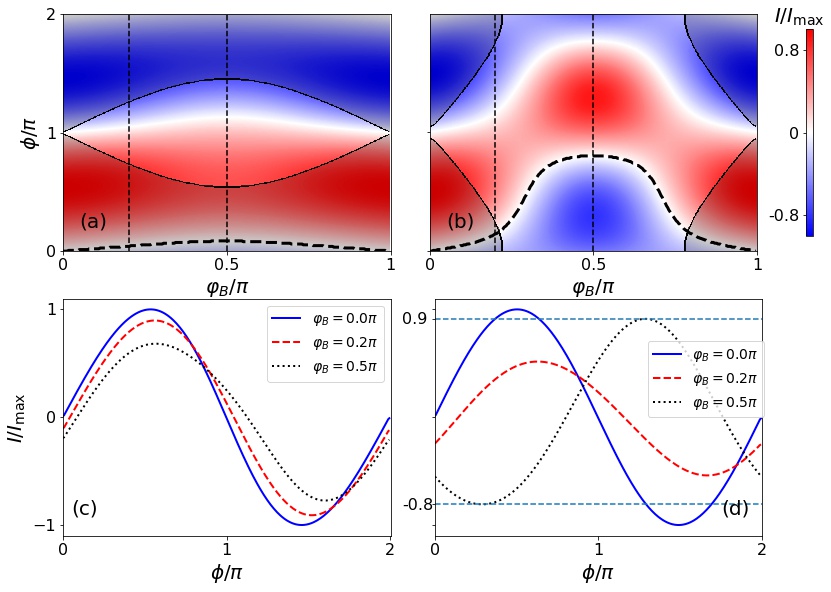}}
\caption{[Color online] [(a) and (b)] Plot of $I(\phi)$ and $Q$ as a 
function of $\phi$ and $\varphi_B$, with $\theta_\textrm{SO} = 
\theta_c=0$, for a HgTe junction with (a) $B=0.4$T and (b) $B=0.8$T. 
The shaded (unshaded) areas have $Q=1$ ($Q=-1$). The TB simulation 
parameters and the junction geometry are the same as in 
Fig.~\ref{fig:I_Q_vs_phi_vs_Bz}. The vertical dashed lines are at 
$\varphi_B=0.2\pi$ and $\varphi_B=0.5\pi$. (c), (d) The current-phase 
relation along the dashed lines in (a) and (b), respectively. In (d), 
note the difference between the maxima of the forward and reverse 
supercurrents, indicated by the horizontal dashed lines for 
$\varphi_B=0.5\pi$ case. }
\label{fig:I_Q_vs_phi_vs_varphiB_BAniso}
\end{figure}

At zero magnetic field the CPR is anti-symmetric under reflection with 
respect to $\phi=\pi$, i.e., $I(\phi)= -I(2\pi-\phi)$. In the presence 
of Rashba SOC and a finite in-plane magnetic field this symmetry is 
preserved only when the field is parallel to the current 
($\varphi_B=0$). The symmetry breaking for $B\neq 0$ and 
$\varphi_B=\pi/2$ is clearly seen in 
Fig.~\ref{fig:I_Q_vs_phi_vs_Bz}(a).

The two white traces in Fig.~\ref{fig:I_Q_vs_phi_vs_Bz}(a) correspond 
to parameters for which the supercurrent vanishes. However, the 
conditions in Eq.~(\ref{i-phi-gs}) are satisfied only along the lower 
trace. As shown in the figure, the path of the lower trace is in good 
agreement with the magnetic field dependence of the ground-state phase 
(black dashed line). This illustrates how the magnetic field dependence 
of the CPR in a phase-biased junction can be used to extract the 
ground-state phase of the phase-unbiased junction. The method works 
well when only a single trace in the $B$ dependence of the CPR 
satisfies Eq.~(\ref{i-phi-gs}). However, the situation may not be that 
clear when there are multiple traces obeying Eq.~(\ref{i-phi-gs}). In 
such a case, different traces correspond to different free energy 
minima, and the information in the $B$ dependence of the CPR is not 
enough to decide which one represents the absolute minimum.

The phase dependence of the energy spectrum is shown in 
Figs.~\ref{fig:I_Q_vs_phi_vs_Bz}(b) and \ref{fig:I_Q_vs_phi_vs_Bz}(c) 
for $B=0.4$~T and $B=0.8$~T [indicated with vertical dashed lines in 
(a)], respectively. The red lines illustrate the formation of MBSs in 
the TS state. For $B=0.4$~T, zero-energy MBSs appear in a reduced 
interval of $\phi$-values, while at $B=0.8$~T zero-energy MBSs exist 
for any value of the phase difference. This is an agreement with the 
topological region depicted in Figs.~\ref{fig:I_Q_vs_phi_vs_Bz}(a). 
Although the junction is long enough, and the MBSs are typically well 
separated from each other [note that MBSs have zero energy in most of 
the topological region, as shown in Figs.~\ref{fig:I_Q_vs_phi_vs_Bz} 
(b) and (c)], they start to hybridize and depart from zero energy when 
the topological gap becomes too small. This is particularly noticeable 
in Fig.~\ref{fig:I_Q_vs_phi_vs_Bz}(c) around $\phi \sim 1.5\pi$, and 
emphasizes the fact that, even if the system is in the topological 
state, one may still need to optimize the topological gap to realize 
stable MBSs.

The Rashba SOC is rotationally invariant, however its combination with 
the in-plane Zeeman interaction leads to magnetoanisotropic effects. 
The magnetic anisotropy of the CPR is shown if 
Figs.~\ref{fig:I_Q_vs_phi_vs_varphiB_BAniso}(a) and 
\ref{fig:I_Q_vs_phi_vs_varphiB_BAniso}(b), where the normalized 
supercurrent is shown as a function of the phase difference, $\phi$, 
and the magnetic field orientation, $\varphi_B$ for the two values of 
magnetic field amplitudes corresponding to the vertical dashed lines in 
Fig.~\ref{fig:I_Q_vs_phi_vs_Bz}(a). In both cases, the topological 
region exhibits a strong dependence on $\varphi_B$. When the in-plane 
magnetic field is parallel to the supercurrent direction, the system is 
in the trivial state for any phase difference. As the magnetic field is 
rotated towards the junction direction, the range of phase differences 
leading to TS increases. The results demonstrate the convenience of 
orienting the magnetic field in the direction, $\varphi_B = \pi/2$, 
i.e., perpendicular to the supercurrent 
flow.~\cite{Scharf2019:PRB,Pakizer2021:PRR} The CPR at different 
magnetic field orientations indicated by vertical dashed lines in 
Figs.~\ref{fig:I_Q_vs_phi_vs_varphiB_BAniso}(a) and \ref{fig:I_Q_vs_phi_vs_varphiB_BAniso}(b) are shown in 
Figs.~\ref{fig:I_Q_vs_phi_vs_varphiB_BAniso}(c) and \ref{fig:I_Q_vs_phi_vs_varphiB_BAniso}(d), respectively. In both cases, the CPR shows a slight 
deviation from a sinusoidal function, evidencing the low transparency 
of the junction. Furthermore, an anomalous phase~\cite{Dolcini2015:PRB, 
Nesterov2016:PRB,Baumard2020:PRB,Zazunov2009:PRL, Yokoyama2014:PRB, 
Hoshino2018:PRB} emerges due to the combined action of the Rashba SOC 
and the in-plane magnetic field, producing a $\varphi_B$-dependent 
shift of the CPR. Although hardly notable at the scale of the figure, 
numerical evaluation reveals that for $\varphi_B\neq 0$, the forward 
(maximum) and reverse (minimum) supercurrents are slightly different. 
This manifestation of the superconducting spin diode 
effect\cite{Ando2020:N} is perhaps more apparent in the CPR shown in 
Fig.~\ref{fig:I_Q_vs_phi_vs_varphiB_BAniso}(d) for $B=0.8$~T and 
$\varphi_B=\pi/2$, where the horizontal lines indicate the different 
amplitudes of the forward and reverse supercurrents.

\subsection{Effects of crystalline anisotropy}

The linear Rashba SOC exhibits rotational invariance. However, the 
coexistence of Rashba and Dresselhaus SOCs reduce the symmetry of the 
spin-orbit field to a twofold $C_{2v}$ symmetry~\cite{Zutic2004:RMP, 
Fabian2007:APS}. Such a symmetry reduction leads to various 
magnetoanisotropic phenomena in both the normal~\cite{Moser2007:PRL, 
Badalyan2009:PRB, Matos-Abiague2015:PRL,Gmitra2013:PRL} and 
superconducting~\cite{Ikegaya2017:PRB, Biderang2018:PRB, Hogl2015:PRL, 
Costa2019:PRB, Alidoust2021:PRB,Martinez2020:PRApp} states as well as 
crystalline anisotropic phenomena in which the system properties depend 
on the specific crystallographic orientation and/or transport 
direction~\cite{Hupfauer2015:NC, Pakizer2021:PRR, Rushforth2007:PRL}. 
The effects of crystalline anisotropy on the topological gap of planar 
JJs was investigated in Ref.~\onlinecite{Pakizer2021:PRR}. In this 
Section we explore the SOC-induced crystalline anisotropy of the CPR in 
a phase-biased, planar JJ. For the numerical simulations of the 
crystalline anisotropy, we consider Al/InSb JJs, which for the 
considered system parameters (see 
Appendix~\ref{SECT:APP:TB_simulations}) have an estimated zero-field 
superconducting coherence length, $\xi=164$nm, which is larger than 
$W_N$, but shorter than $W_S$.

\begin{figure}[t]
\centerline{\includegraphics[width=1.05\columnwidth]{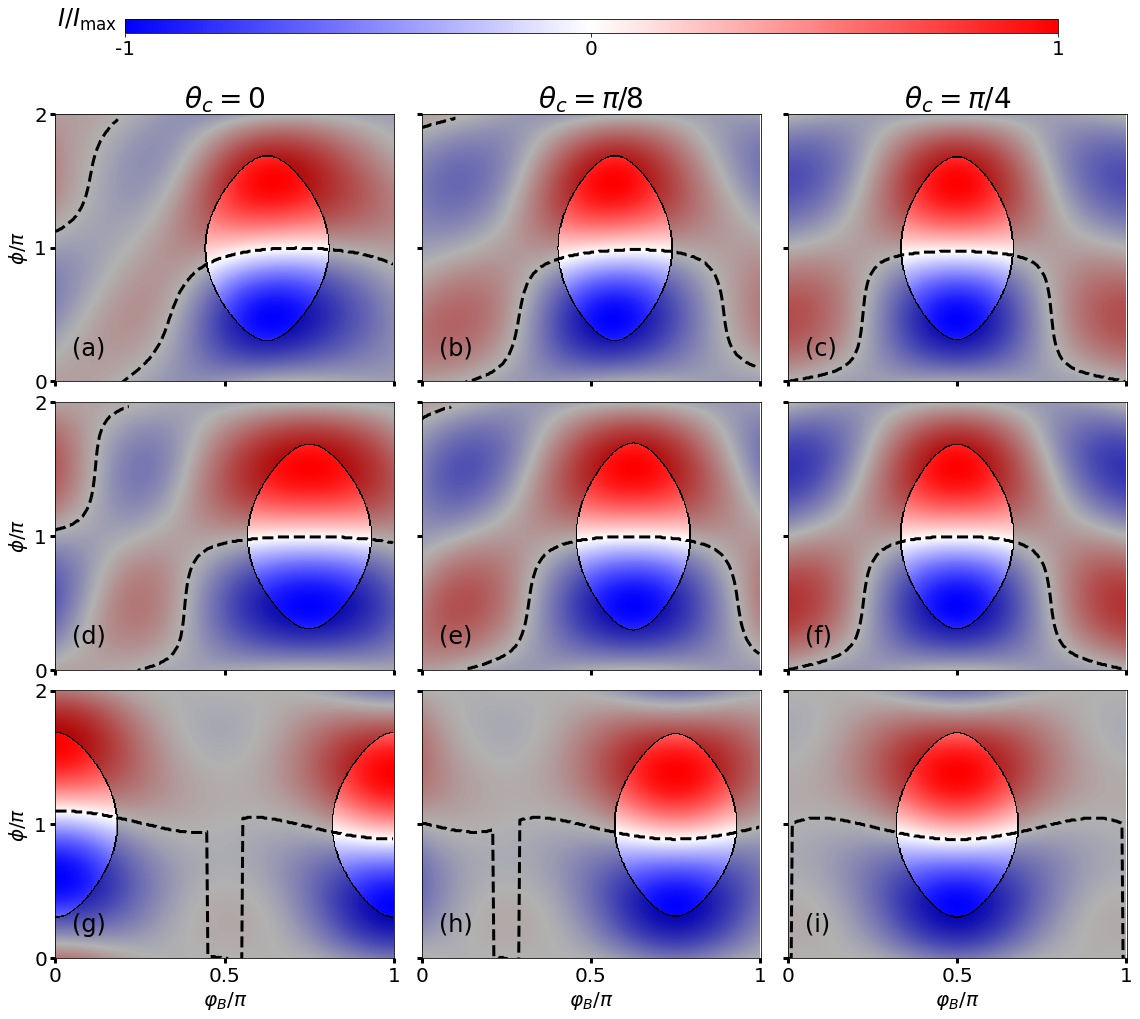}}
\caption{[Color online] [(a)--(i)] Plot of $I(\phi)$  and $Q$ as a 
function of $\phi$ and $\varphi_B$ for an InSb junction, for various 
values of $\theta_\textrm{SO}$ and $\theta_c$, for $B=0.6$T [see 
Figs.~\ref{fig:I_Q_vs_phi_vs_Bz}(b) and 
\ref{fig:I_Q_vs_phi_vs_varphiB_BAniso}(c)] . The top row has 
$\theta_\textrm{SO}=\pi/8$, the middle row has 
$\theta_\textrm{SO}=\pi/4$ and the bottom row has 
$\theta_\textrm{SO}=\pi/2$. The shaded (unshaded) areas have $Q=1$ 
($Q=-1$). The TB simulation parameters and the junction geometry are 
given in Appendix.}
\label{fig:I_Q_vs_phi_vs_varphiB_xtalAniso}
\end{figure}

The CPR as a function of the in-plane magnetic field direction is shown 
in Fig.~\ref{fig:I_Q_vs_phi_vs_varphiB_xtalAniso} for different 
supercurrent directions ($\theta_c$) and spin-orbit angles 
($\theta_{so}$). The supercurrent direction is fixed by the junction 
orientation with respect to the $[100]$ crystallographic axis. The 
dashed lines indicate the ground-state phase computed by minimizing the 
system free energy and are in good agreement with the CPR contours 
obeying Eq.~(\ref{i-phi-gs}). When only Rashba SOC is present, 
$\theta_{so}=0$ and the CPR is independent of $\theta_c$. However, the 
presence of Dresselhaus SOC leads to appreciable changes in the CPR. 
For $\theta_{so}=\pi/8$, the topological (unshaded) region changes its 
size and exhibits a shift in its position with respect to the magnetic 
field direction when the supercurrent direction changes, as shown in 
Figs.~\ref{fig:I_Q_vs_phi_vs_varphiB_xtalAniso}(a)--\ref{fig:I_Q_vs_phi_vs_varphiB_xtalAniso}(c). 
A similar trend is observed when the strength of the Rashba and 
Dresselahus SOC are equal, $\alpha=\beta$ [i.e., $\theta_{so}=\pi/4$ in 
Figs.~\ref{fig:I_Q_vs_phi_vs_varphiB_xtalAniso}(d) and 
\ref{fig:I_Q_vs_phi_vs_varphiB_xtalAniso}(e)] and when only Dresselhaus 
SOC is present [i.e., $\theta_{so}=\pi/2$ in 
Figs.~\ref{fig:I_Q_vs_phi_vs_varphiB_xtalAniso}(g)--\ref{fig:I_Q_vs_phi_vs_varphiB_xtalAniso}(i)].

Previous investigations~\cite{Pientka2017:PRX, Setiawan2019:PRB2, 
Dartiailh2021:PRL} have shown the importance of properly tuning the 
chemical potential, magnetic field strength, and superconducting phase 
difference for driving the JJ into the TS state.  However, the strong 
dependence of the TS state and the ground-state phase on the spin-orbit 
angle ($\theta_{so}$), the magnetic field orientation ($\varphi_B$), 
and the supercurrent direction ($\theta_c$), shown in 
Fig.~\ref{fig:I_Q_vs_phi_vs_varphiB_xtalAniso}, reveals that an 
experimental setup with an adequate combination of $\theta_{so}$, 
$\varphi_B$, and $\theta_c$ values is also crucial for inducing TS in 
planar JJs.\\

\section{Topological superconductivity in phase-unbiased Josephson junctions}\label{SECT:PhaseUnbiasedJJ}

\subsection{Effects of magnetoanisotropy}

\begin{figure}[t]
\centerline{\includegraphics[width=1.05\columnwidth]{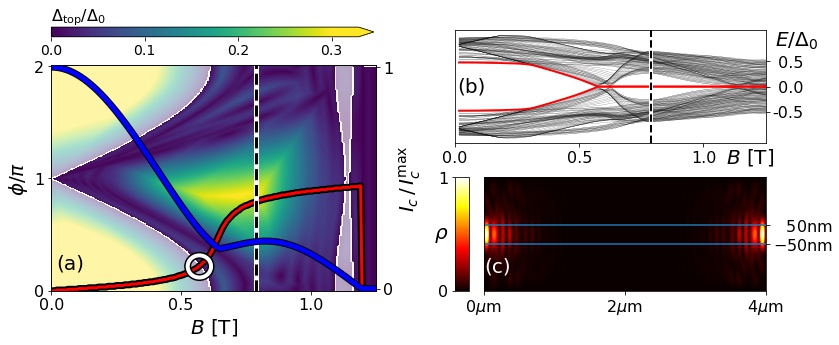}}
\caption{[Color online] (a) Plot of $\Delta_\textrm{top}/\Delta_0$ and 
$Q$ as a function of $\phi$ and $B$ for a HgTe Josephson junction (see 
Fig.~\ref{fig:jj}), with $\theta_\textrm{SO}=\theta_c=0$ and 
$\varphi_B=\pi/2$. The shaded (unshaded) areas have $Q=1$ ($Q=-1$). The 
TB simulation parameters and the junction geometry are the same as in 
Fig.~\ref{fig:I_Q_vs_phi_vs_Bz}. The red lines are the ground state 
phase $\phi_{GS}$ of the system and the blue lines are 
$I_c/I_c^\textrm{max}$, $I_c^\textrm{max}$ being the maximum critical 
current in the junction at $T=0.7$K. The white ring indicates the field 
and phase values at which the topological transition occurs. (b) The 
lowest 200 energy levels calculated along the red ($\phi_{GS}$) curve 
in (a). Here, a fully 2D closed system with $L=4002$nm is used instead 
of the translationally invariant system used in (a). For (a) and (b) 
both, the vertical dashed line marks the $B=0.79$T value at which 
$\Delta_\textrm{top}/\Delta_0$ is largest along the $\phi_{GS}$ curve 
within the nontrivial region. (c) The density plot of $\rho=|\Psi|^2$ 
for the Majorana mode (b) (red curve) at $B=0.79$T. }
\label{fig:Gap_Q_vs_phi_vs_Bz_phiGS}
\end{figure}

In the absence of a magnetic flux, the superconducting phase difference 
self-tunes to a value (the ground-state phase, $\phi_{GS}$) that leads 
to the minimization of the system free energy. Since the free energy of 
the junction varies with the applied magnetic field, both the 
ground-state phase and the critical current also change as the strength 
of the in-plane magnetic field increases.

The topological gap [see Eq.~(\ref{topogap})] of a HgTe JJ (where only 
Rashba SOC is relevant) with an in-plane magnetic field perpendicular 
to the supercurrent direction is shown in 
Fig.~\ref{fig:Gap_Q_vs_phi_vs_Bz_phiGS}(a) as a function of the phase 
difference and magnetic field amplitude. The shaded and unshaded 
regions correspond to trivial ($Q=1$) and topological ($Q=-1$) phases. 
The red and blue lines correspond to the ground-state phase and 
normalized critical current, respectively. The ground-state phase 
exhibits a jump from 0 to $\pi$ as the magnetic field increases, as 
shown in Fig.~\ref{fig:Gap_Q_vs_phi_vs_Bz_phiGS}(a) and earlier in 
Fig.~\ref{fig:I_Q_vs_phi_vs_Bz}(a). The junction transitions into the 
topological state at the center of the white ring, where $\phi_{GS}$ 
crosses the topological region. The ground-state phase jump is 
accompanied by a local minimum in the critical current (blue line). 
However, due to the smoothness of the ground-state jump, the critical 
current minimum may occur at a magnetic field higher than the 
transition field (corresponding to the center of the white ring). 
Therefore, in the best situation, the critical current minimum alone 
can only be an indirect indication of the topological phase transition. 
In a more general scenario, there are situations in which the 
topological transition occurs without the current having a local 
minimum~\cite{Pakizer2021:PRB,Setiawan2019:PRB2}.

Figure \ref{fig:Gap_Q_vs_phi_vs_Bz_phiGS}(a) reveals that the 
topological gap is sizable only on a reduced part of the topological 
region. The topological gap is crucial for ensuring the practical 
protection of the MBSs. Therefore, only the portion of the parameter 
space leading to the TS state with a sizable topological gap is useful 
from a practical point of view. The departure of $\phi_{GS}$ from the 
value 0 as the magnetic field increases, yields a topological 
transition at a magnetic field slightly smaller than the one required 
at $\phi=0$. Furthermore, the self-tuned jump of the ground-state phase 
to values close to $\pi$ allows for achieving a sizable topological gap 
in the TS state.

The ground-state spectrum, i.e., the energy spectrum for which the 
junction has the minimum free energy, is shown in 
Fig.~\ref{fig:Gap_Q_vs_phi_vs_Bz_phiGS}(b) as a function of the 
magnetic field strength. At zero magnetic field the system is in the 
trivial state with a gap of about $\Delta_0/2$. As the magnetic field 
increases, the ground-state phase starts to depart from zero, yielding  
a decrease in the energy gap until it closes and reopens at the 
topological transition. Once the system enters the TS state, zero 
energy MBSs (red line) emerge inside the topological gap. The 
probability density of the MBSs (normalized to its maximum value) is 
shown in Fig.~\ref{fig:Gap_Q_vs_phi_vs_Bz_phiGS}(c) for a magnetic 
field value indicated by the vertical dashed line in (b). The MBSs are 
well localized at the ends of the junction.

\begin{figure}[t]
\centerline{\includegraphics[width=0.8\columnwidth]{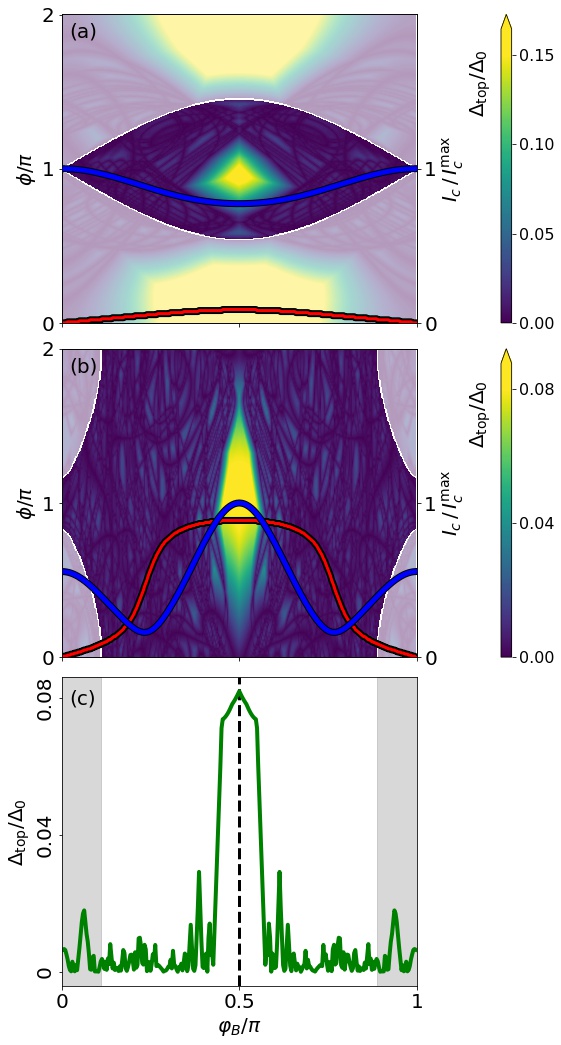}}
\caption{[Color online] [(a) and (b)] Plot of 
$\Delta_\textrm{top}/\Delta_0$ and $Q$ as a function of $\phi$ and 
$\varphi_B$ for a HgTe junction, for (a) $B=0.4$T and (b) $B=1.0$T. The 
shaded (unshaded) areas have $Q=1$ ($Q=-1$). The TB simulation 
parameters and the junction geometry are the same as in 
Fig.~\ref{fig:I_Q_vs_phi_vs_Bz}. The red lines are the ground state 
phase $\phi_{GS}$ of the system and the blue lines are 
$I_c/I_c^\textrm{max}$, $I_c^\textrm{max}$ being the maximum critical 
current in the junction for the respective $B$ value and at $T=0.7$K. 
(c) Plot of $\Delta_\textrm{top}$ along $\phi=\phi_{GS}$ curve (red 
line) in (b) as a function of $\varphi_B$. The shaded (unshaded) areas 
have $Q=1$ ($Q=-1$) at the given $\varphi_B$. The vertical dashed line 
corresponds to $\varphi_B^\textrm{opt}$ given by Eq.~(\ref{opt-mag}). 
$\theta_\textrm{SO}=\theta_c=0$ for all plots.}
\label{fig:Gap_Q_vs_phi_vs_varphiB_GapAtphiGS_BAniso}
\end{figure}

The magnetic anisotropy of the topological gap, the ground-state phase 
(red line), and the critical current is shown in 
Fig.~\ref{fig:Gap_Q_vs_phi_vs_varphiB_GapAtphiGS_BAniso}(a) for 
$B=0.4$~T. The trajectory of the ground-state phase indicates that for 
such a field value the system is unable to self-tune into the TS state 
for any magnetic field orientation. The critical current (blue line) 
exhibits a local minimum but it is not associated to a topological 
transition. However, if the field amplitude is increased to 1~T, the 
self-tuning of the ground-state phase can drive the system into the TS 
state when the magnetic field is rotated away from the direction of the 
supercurrent, as shown in 
Fig.~\ref{fig:Gap_Q_vs_phi_vs_varphiB_GapAtphiGS_BAniso}(b). In this 
case, the jump in the ground-state phase not only allows for the 
topological transition, but also for a finite topological gap when 
$\varphi_B =\pi/2$. The jumps in the ground-state phase are accompanied 
by critical current minima. Interestingly, the critical current 
exhibits a maximum at the magnetic orientation leading to the TS state 
with the largest topological gap.

The ground-state topological gap (i.e., the topological gap at the 
ground-state phase) as a function of the magnetic field orientation for 
a phase-unbiased Al/HgTe JJ is represented by the green line in 
Fig.~\ref{fig:Gap_Q_vs_phi_vs_varphiB_GapAtphiGS_BAniso}(c). As the 
in-plane magnetic field is rotated from a direction parallel 
($\varphi_B=0$) to perpendicular ($\varphi_B=\pi/2$) to the 
supercurrent direction, the topological gap self-tunes from a 
negligible small value to a maximum of about $0.08\Delta_0$. The figure 
illustrates the importance of properly orienting the magnetic field 
when driving the system into a robust TS state, and evidences that even 
if the ground-state phase self-tuning can drive the system in the TS 
state for a wide range of magnetic field orientations (white area), the 
topological gap is sizable and stable only within a small window around 
$\varphi_B=\pi/2$. This is consistent with recent experimental results, 
where the TS state deteriorates as the magnetic field deviates from the 
direction perpendicular to the supercurrent.~\cite{Dartiailh2021:PRL}

\subsection{Effects of crystalline anisotropy}

\begin{figure}[t]
\centerline{\includegraphics[width=1.05\columnwidth]{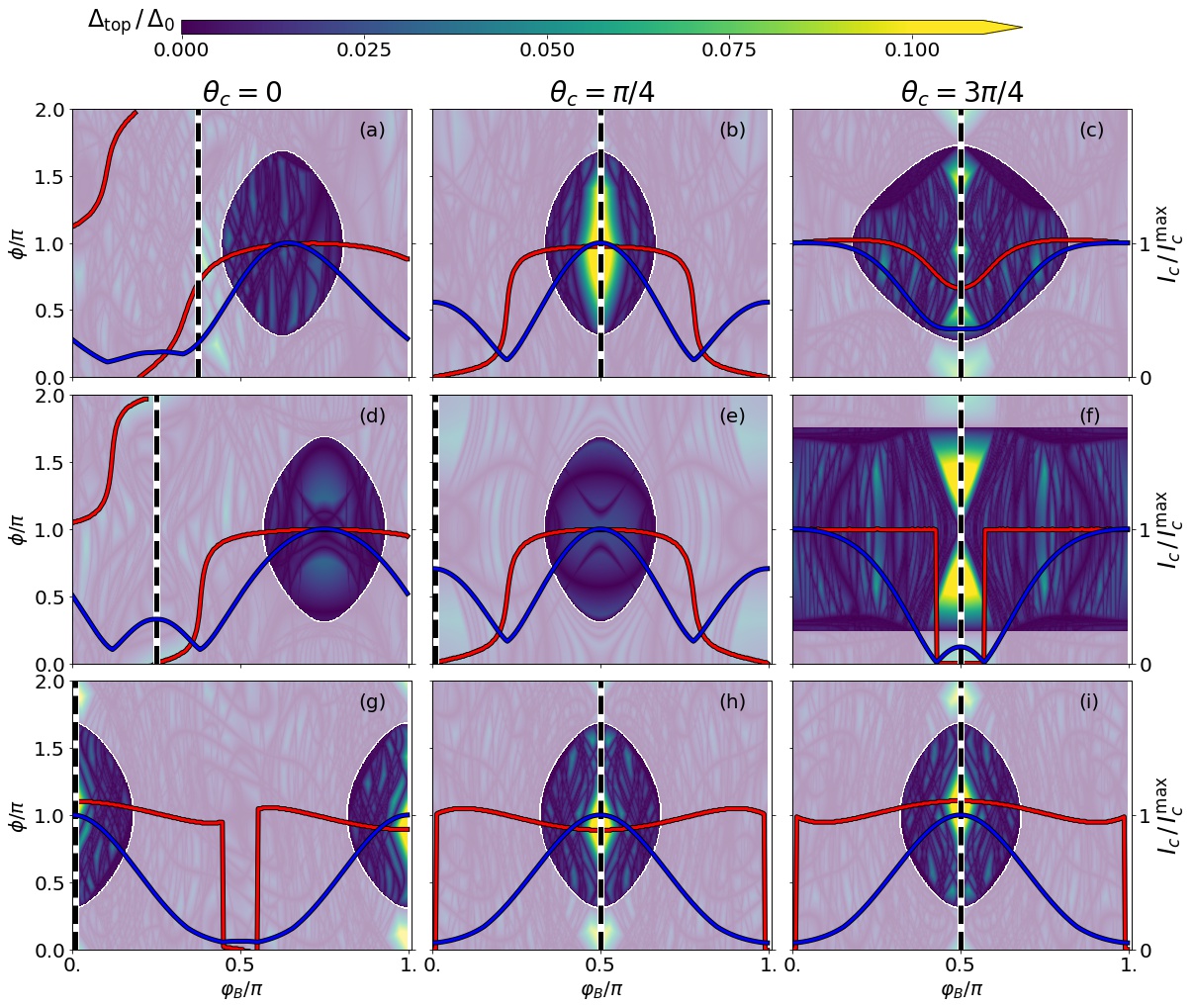}}
\caption{[Color online] (a)-(i) Plot of $\Delta_\textrm{top}/\Delta_0$  
and $Q$ as a function of $\phi$ and $\varphi_B$ for an InSb junction, 
for various values of $\theta_\textrm{SO}$ and $\theta_c$. The top row 
has $\theta_\textrm{SO}=\pi/8$, the middle row has 
$\theta_\textrm{SO}=\pi/4$ and the bottom row has 
$\theta_\textrm{SO}=\pi/2$. For all of the figures, $B=0.6$T. The 
shaded (unshaded) areas have $Q=1$ ($Q=-1$). The TB simulation 
parameters and the junction geometry are specified in the Appendix. The 
red lines are the ground state phase $\phi_{GS}$ of the system and the 
blue lines are $I_c/I_c^\textrm{max}$, $I_c^\textrm{max}$ being the 
maximum critical current in the junction at $T=0.7$K.  The vertical 
dashed lines correspond to $\varphi_B^\textrm{opt}$ given by 
Eq.~(\ref{opt-mag}), for the respective $\theta_\textrm{SO}$ and 
$\theta_c$. }
\label{fig:Gap_Q_vs_phi_vs_varphiB_xtalAniso}
\end{figure}

\begin{figure}[t]
\centerline{\includegraphics[width=1.05\columnwidth]{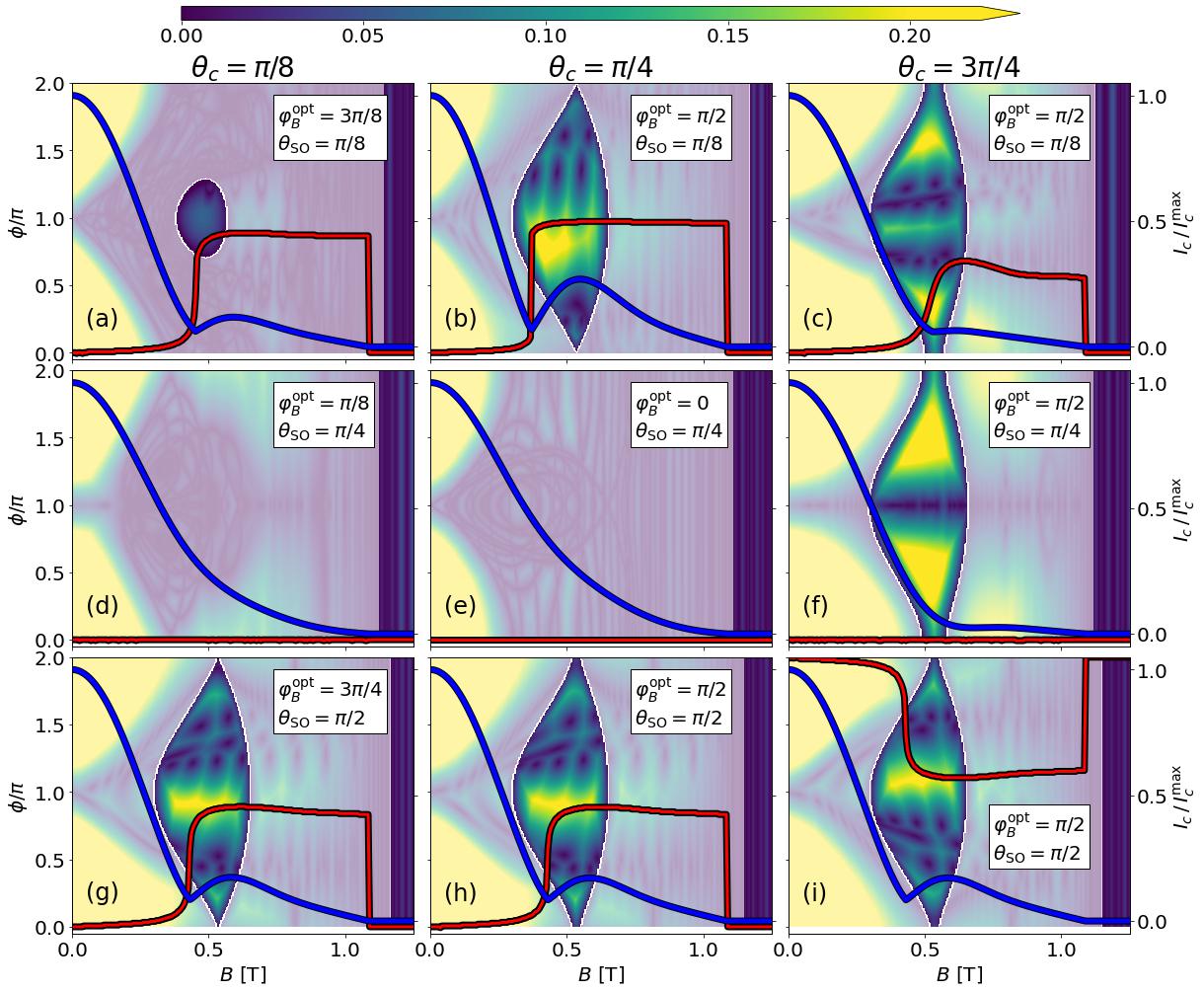}}
\caption{[Color online] (a)-(i) Plot of $\Delta_\textrm{top}/\Delta_0$  
and $Q$ as a function of $\phi$ and $B$ for an InSb junction, for 
various values of $\theta_\textrm{SO}$ and $\theta_c$. The shaded 
(unshaded) areas have $Q=1$ ($Q=-1$). The TB simulation parameters and 
the junction geometry are provided in the Appendix. The red lines are 
the ground state phase $\phi_{GS}$ of the system and the blue lines are 
$I_c/I_c^\textrm{max}$, $I_c^\textrm{max}$ being the maximum critical 
current in the junction at $T=0.7$K. For each plot, $\varphi_B = 
\varphi_B^\textrm{opt}$ for the respective $\theta_\textrm{SO}$ and 
$\theta_c$, corresponding to (a) $\varphi_B=3\pi/8$, (b) 
$\varphi_B=\pi/2$, (c) $\varphi_B=\pi/2$, (d) $\varphi_B=\pi/8$, (e) 
$\varphi_B=0$, (f) $\varphi_B=\pi/2$, (g) $\varphi_B=3\pi/4$, (h) 
$\varphi_B=\pi/2$ and (i) $\varphi_B=\pi/2$. }
\label{fig:Gap_Q_vs_phi_vs_Bz_optVarphiB_xtalAniso}
\end{figure}

To explore the effects of crystalline anisotropy on phase-unbiased JJs, 
we consider an Al/InSb junction (see Appendix 
\ref{SECT:APP:TB_simulations} for system parameters), where Rashba and 
Dresselhaus SOCs coexist. The topological gap as a function of the 
magnetic field orientation is shown in 
Fig.~\ref{fig:Gap_Q_vs_phi_vs_varphiB_xtalAniso} for $B=0.6$~T, 
different crystallographic orientations ($\theta_c$) of the junction, 
and various values of the spin-orbit angle ($\theta_{so}$). The top row 
corresponds to $\theta_{so}=\pi/8$, i.e., to a situation in which 
Rashba SOC is about 2.4 times stronger than Dresselhaus SOC. The middle 
row displays the case $\theta_{so}=\pi/4$, in which Rashba and 
Dressellhaus SOCs have equal strength, and the bottom row with 
$\theta_{so}=\pi/2$ corresponds to a junction in which only Dresselhaus 
SOC is present. The case of junctions in which only Rashba SOC is 
present has been omitted because such junctions do not exhibit 
crystalline anisotropy. Shaded and unshaded regions represent trivial 
($Q=1$) and topological ($Q=-1$) states, respectively. The blue lines 
represent the normalized positive branch of the critical current, while 
the red lines indicate the ground-state phase behavior. The topological 
regions, critical current, and ground-state phase exhibit a strong 
dependence on both the magnetic field and junction orientations. 
Furthermore, the critical current dependence on the magnetic field 
direction can be used to determine whether both Rashba and Dresselhaus 
or only one SOC interaction is present in the system. Indeed, as long 
as the junction is not oriented along the spin-orbit field symmetry 
axes [i.e., $\theta_c\neq (2n+1)\pi/4$] and only Rashba, only 
Dresselhaus, or both SOCs are present, the absolute maxima of the 
critical current occur for magnetic field orientations 
$\varphi_B=(2n+1)\pi/2$ [see 
Fig.~\ref{fig:Gap_Q_vs_phi_vs_varphiB_GapAtphiGS_BAniso}(b)], 
$\varphi_B = n\pi$ [see Fig. 
~\ref{fig:Gap_Q_vs_phi_vs_varphiB_xtalAniso}(g)], or $\varphi_B\neq 
n\pi/2$ (with $n$ being an integer number) [see 
Figs.~\ref{fig:Gap_Q_vs_phi_vs_varphiB_xtalAniso}(a) and 
\ref{fig:Gap_Q_vs_phi_vs_varphiB_xtalAniso}(d)], respectively.

The spin-orbit field in zinc-blende semiconductor quantum wells grown 
along the $[001]$ crystallographic direction exhibits a $C_{2v}$ 
symmetry, with symmetry axes along the $[110]$ and $[\bar{1}10]$ 
directions.~\cite{Fabian2007:APS} Although the specific crystallographic 
direction of the junction may lower the symmetry to $C_2$, the $C_{2v}$ 
symmetry is still preserved as long as the junction direction coincides 
with one of the spin-orbit field symmetry axes (i.e., when 
$\theta_c=(2n+1)\pi/4$, with $n$ being an integer number). This is the 
situation in the middle and right columns in 
Figs.~\ref{fig:Gap_Q_vs_phi_vs_varphiB_xtalAniso}, where the 
topological gap, ground-state phase, and critical current exhibit a 
$C_{2v}$ symmetry with respect to the magnetic field orientation with a 
symmetry axis $\varphi_B=\pi/2$. Note that for $\theta_c=\pi/4$ and 
$\theta_c=3\pi/4$, $\varphi_B=\pi/2$ corresponds to magnetic fields 
along the $[\bar{1}10]$ and $[\bar{1}\bar{1}0]$ directions, 
respectively (i.e., to magnetic field directions along symmetry axes of 
the spin-orbit field).

Unlike the critical current dependence on the magnetic field strength, 
which may exhibit minima (accompanied by ground-state phase jumps) when 
the system transits from the trivial to the topological superconducting 
state~\cite{Pientka2017:PRX,Dartiailh2021:PRL} (see also 
Fig.~\ref{fig:Gap_Q_vs_phi_vs_Bz_optVarphiB_xtalAniso}), the minima of 
the critical current dependence on the magnetic field direction (see 
Fig.~\ref{fig:Gap_Q_vs_phi_vs_varphiB_xtalAniso}) is not an indicator 
of topological phase transitions.

The results shown in Fig.~\ref{fig:Gap_Q_vs_phi_vs_varphiB_xtalAniso} 
reveal that the realization of the topological superconducting state 
with a sizable topological gap requires an adequate orientation of the 
magnetic field, according to the junction crystallographic direction. 
Indeed, even if the system is in the topological state, the topological 
gap protecting the MBSs can be very small when the optimal magnetic 
field orientation (vertical, dashed lines) calculated from 
Eq.~(\ref{opt-mag}) do not cross the topological region, as shown in 
Figs.~\ref{fig:Gap_Q_vs_phi_vs_varphiB_xtalAniso}(a), 
\ref{fig:Gap_Q_vs_phi_vs_varphiB_xtalAniso}(d), and 
\ref{fig:Gap_Q_vs_phi_vs_varphiB_xtalAniso}(e). However, a sizable 
topological gap is achieved when the system is in the TS state and the 
magnetic field orientation fulfills Eq.(\ref{opt-mag}), as illustrated 
in Figs.~\ref{fig:Gap_Q_vs_phi_vs_varphiB_xtalAniso}(b), 
\ref{fig:Gap_Q_vs_phi_vs_varphiB_xtalAniso}(c), and 
\ref{fig:Gap_Q_vs_phi_vs_varphiB_xtalAniso}(f)--\ref{fig:Gap_Q_vs_phi_vs_varphiB_xtalAniso}(i).

The topological gap as a function of the magnetic field strength is 
shown in Fig.~\ref{fig:Gap_Q_vs_phi_vs_Bz_optVarphiB_xtalAniso} for for 
the corresponding optimal magnetic field orientations ($\varphi_B$), 
different crystallographic orientations ($\theta_c$) of the junction, 
and various values of the spin-orbit angle ($\theta_{so}$). For 
$\theta_{so}=\pi/8$ and $\theta_{so}=\pi/2$ (upper and lower rows, 
respectively), as the magnetic field strength is increased, the system 
transits into the TS state when the ground-state phase (red line) jumps 
and crosses into the topological region (unshaded zones). The proper 
orientation of the magnetic field allows for the existence of a sizable 
topological gap when the system enters in the TS state. The jump of the 
ground-state phase is accompanied by a minimum in the critical current. 
Such a behavior has previously been used as a signature of TS phase 
transitions.~\cite{Pientka2017:PRX,Dartiailh2021:PRL} Note, however, 
that in junctions with narrow S regions critical current minima may not 
necessarily signal topological 
transitions~\cite{Setiawan2019:PRB,Pakizer2021:PRB} and the observation 
of more reliable signatures in other physical quantities, such as the 
spin susceptibility might be required~\cite{Pakizer2021:PRB}. When the 
Rashba and Dresselhaus SOCs have equal strengths (i.e., 
$\theta_{so}=\pi/4$) and $\theta_c=\pi/8$ or $\theta_c=\pi/4$, the 
system remains in the trivial state for all magnetic field strengths 
and no ground-state jump nor local critical current minimum occur. 
However, the system can still reach the TS state when 
$\theta_c=3\pi/4$, even without a ground-state phase jump (nor 
associated critical current minimum). However, the absence of the phase 
jump does not allow the system to reach the TS state with an optimal 
topological gap. The results suggest that coexisting Rashba and 
Dresselhaus SOCs with equal strength may not be favorable for realizing 
stable TS in phase-unbiased JJs.

\section{Summary}

We show that a rich interplay of magnetic field direction, crystalline 
orientation, and the relative strengths between Rashba and Dresselhaus 
SOCs (parametrized by a spin-orbit angle), all play an important role 
in defining the optimal set of parameters that lead to a large 
topological gap in a topologically nontrivial state. We provide 
examples of the effects of magnetoanisotropy on the current-phase 
relation of a proximitized Al/HgTe planar JJ, where only Rashba SOC is 
sizable, and show that both the topological phase diagram and the CPR 
strongly depend on the direction of the applied magnetic field. We 
demonstrate that for a phase-unbiased planar JJ in which the phase is 
allowed to self-tune to its ground-state value (i.e., the phase value 
that minimizes the system free energy), changing the magnetic field 
direction can cause $\pi$-jumps in the ground-state phase. We also 
consider the case of an Al/InSb junction, where both Rashba and 
Dresselhaus SOCs are relevant and not only the magnetic field 
direction, but the junction crystallographic orientation and spin-orbit 
angle also affect the topological superconducting state. We show that 
to realize stable MBSs, the system parameters must be tuned such that 
the junction is in the topological superconducting state and a sizable 
topological gap is achieved by properly orienting the magnetic field, 
in dependence of the junction crystallographic orientation and the 
spin-orbit angle parametraizing the relative strengths between Rashba 
and Dresselhaus SOCs. In phase-unbiased JJs the ground-state phase 
self-tunes and can exhibit $\pi$-jumps as the magnetic field is 
rotated. When the magnetic field orientation is optimal and the field 
strength is varied, the ground-state selftuning $\pi$-jumps enables the 
system transition to the topological superconducting state with a 
sizable topological gap.

\noindent\emph{Acknowledgments.} This work was supported by DARPA Grant 
No. DP18AP900007.


%%%%%%%%%%%%%%%%%%%%%%%%%%%%%%%%%%%%%%%%%%%%%%%%%%%%%%%%%%%%%
\appendix
%%%%%%%%%%%%%%%%%%%%%%%%%%%%%%%%%%%%%%%%%%%%%%%%%%%%%%%%%%%%%
\section{Tight-binding simulations}\label{SECT:APP:TB_simulations}

\begin{figure}[t]
\centerline{\includegraphics[width=0.9\columnwidth]{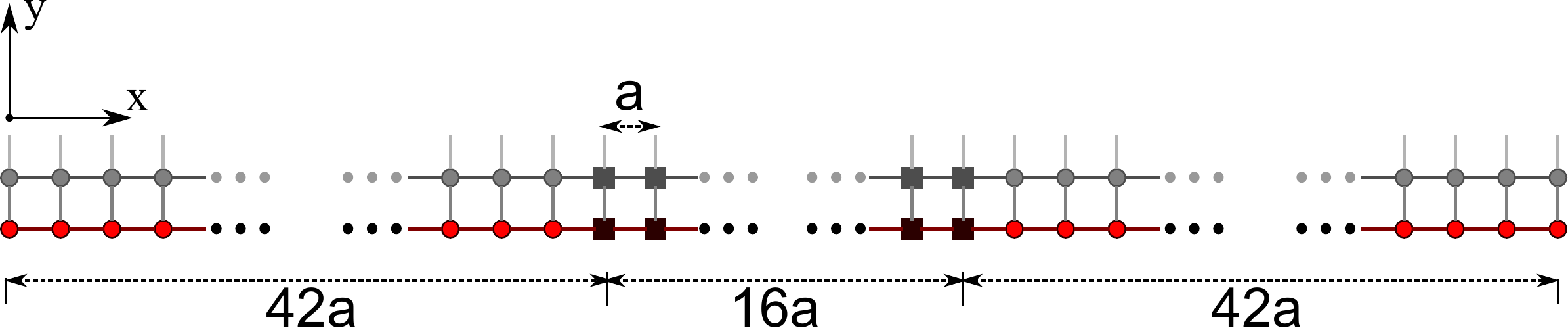}} 
\caption{[Color online] The tight-binding simulation 
lattice with the lattice constant $a$. The square sites correspond to 
the Josephson junction and the round sites correspond to the 
superconducting leads. The grayed out copies of the system represent 
the translational invariance in the $y$-direction of the system. The 
width of the SC leads are $42a=252$nm each and the width of the 
Josephson junction is $16a=96$nm. }
\label{fig:TB_Drawing}
\end{figure}

In this Appendix, we briefly describe the tight-binding (TB) simulation 
methods used in the main text to obtain 
Figures~\ref{fig:I_Q_vs_phi_vs_Bz}-\ref{fig:Gap_Q_vs_phi_vs_Bz_optVarphiB_xtalAniso}. 
We start by discretizing the Hamiltonian in Eqs.~(\ref{H-BdG}) and 
(\ref{Ho}) on a square lattice in the usual manner:~\cite{Datta2007}
\begin{align}\label{EQN:TB_Hamiltonian}
H_\textrm{TB}           &= \hat{H}^\textrm{onsite}+\left(\hat{V}^\textrm{up}+\hat{V}^\textrm{right} + \textrm{h.c.} \right)\nonumber\\
\hat{H}^\textrm{onsite} &= \sum_{j\geq 0, i} h^\textrm{onsite}(x_i) \, \ket{x_i, y_j}\bra{x_i, y_j} \nonumber\\
\hat{V}^\textrm{up}     &= \sum_{j\geq 0, i} v^\textrm{up}(x_i) \, \ket{x_i, y_j+a}\bra{x_i, y_j} \nonumber\\
\hat{V}^\textrm{right}  &= \sum_{j\geq 0, i} v^\textrm{right}(x_i) \, \ket{x_i+a, y_j}\bra{x_i, y_j},
\end{align}
where $a$ is the lattice constant and $(x_i, y_j) = (i\, a, j\,a)$ are 
the $x-$ and $y-$ coordinates of the lattice points and $i, j$ are 
integers representing the $i^\mathrm{th}$ ($j^\mathrm{th}$) lattice 
point along the $x$- ($y$-) axis. The onsite ($h^\textrm{onsite}$) and 
hopping ($v^\textrm{up}, v^\textrm{right}$) terms do not have 
$y_j$-dependence because we consider a system with translational 
invariance in the $y$-direction, except in the cases where we consider 
the energy levels or the wavefunction amplitude of a 2D closed system 
[see Figs.~\ref{fig:I_Q_vs_phi_vs_Bz}(b) and 
\ref{fig:I_Q_vs_phi_vs_Bz}(c), and 
Figs.~\ref{fig:Gap_Q_vs_phi_vs_Bz_phiGS}(b) and 
\ref{fig:Gap_Q_vs_phi_vs_Bz_phiGS}(c)]. These terms are given by
\begin{align}\label{EQN:TB_terms}
h^\textrm{onsite}(x_i)  &= \left(4t +\epsilon(\lambda, \theta_c) - \mu \right)\,\tau_z \sigma_0 + \mathbf{B}\cdot\boldsymbol{\sigma} \nonumber\\
                        & \qquad +\Delta(x_i) \tau_+ + \Delta^*(x_i) \tau_-,\nonumber\\
v^\textrm{right}(x_i)   &= -t\,\tau_z \sigma_0 +\frac{i}{2a} \alpha \tau_z \sigma_y \nonumber\\
                        & \qquad -\frac{i}{2a}\,\beta \, \left( \cos{2\theta_c} \, \tau_z \sigma_x - \sin{2\theta_c}\, \tau_z \sigma_y\right),\nonumber \\
v^\textrm{up}(x_i)      &= -t\,\tau_z \sigma_0 -\frac{i}{2a} \alpha \tau_z \sigma_x \nonumber\\
                        & \qquad +\frac{i}{2a}\,\beta \, \left( \cos{2\theta_c} \, \tau_z \sigma_y + \sin{2\theta_c}\, \tau_z \sigma_x\right).
\end{align}
Here, $t=\hbar^2/2 m^* a^2$ is the hopping parameter, $m^*$ is the 
effective mass and the definitions of $\tau$, $\sigma$, $\Delta$, 
$\lambda$, $\alpha$, $\beta$, and $\theta_c$ are given in the main text. 
$\epsilon(\lambda, \theta_c) = (2 m^*\lambda^2/\hbar^2) \, 
\left(1+(\sin{2\theta_c})^2 \right)$ is the minimum single particle 
energy, which is at least an order smaller than other relevant energies 
in the systems we consider. An illustration of the lattice 
discretization used for the numerical calculations is shown in 
Fig.~\ref{fig:TB_Drawing}.

In order to calculate the free energy in Eq.~(\ref{EQN:free-energy}) as 
well as the topological gap, we need the energy spectrum of the system. 
The translational invariance in the $y$-direction implies the momentum 
$\hbar k_y$ is a good quantum number. We utilize the Kwant 
package~\cite{Groth2014:NJP} to solve the relevant eigenvalue problem 
for a given $k_y$:
\begin{align}\label{EQN:Kwant_TransInv_Spectrum}
\left(\hat{V}^\textrm{down}_{y=0}\, e^{i\,k_y a} + \hat{H}_0^{y=0} + \hat{V}^\textrm{up}_{y=0}\, e^{-i\,k_y a} \right)\,\psi &=E(k_y) \, \psi
\end{align}
where 
\begin{align}\label{EQN:Kwant_TransInv_Spectrum_IndividualTerms}
\hat{H}_0^{y=0} &= \sum_{i} \left[ h^\textrm{onsite}(x_i) \ket{x_i, y=0}\bra{x_i, y=0} \right.\nonumber\\
				&\qquad \left. + \left( v^\textrm{right}(x_i) \ket{x_i+a, y=0}\bra{x_i, y=0} + \mathrm{h.c.}\right) \right], \nonumber\\
\hat{V}^\textrm{up}_{y=0} &= \sum_{i} v^\textrm{up}(x_i) \, \ket{x_i, y=a}\bra{x_i, y=0} \textrm{ and}\nonumber\\
\hat{V}^\textrm{down}_{y=0} &= \left(\hat{V}^\textrm{up}_{y=0}\right)^{\dagger}.
\end{align}

Finally, to obtain the topological charge $Q$, we make use of the 
formula~\cite{Tewari2012:PRL}
\begin{align}\label{EQN:Topocharge_TB}
Q &= \mathrm{sgn}\left[\frac{\textrm{Pf}\left(H(k_y=\pi/a)\right)}{\textrm{Pf}\left(H(k_y=0)\right)}\right],
\end{align}
where $\textrm{Pf}(.)$ is the Pfaffian and $H(k_y=\pi/a)$, $H(k_y=0)$ 
are obtained using the Kwant package.~\cite{Groth2014:NJP, 
Wimmer2021:ACMTMS} 

The system parameters used in the numerical simulations are summarized 
in Table \ref{TBL:Parameters}.

\begin{table}[ht!] 
\caption{Material and simulation properties of the planar JJs, used 
throughout this work. Here, $m_0$ is the rest mass of the electron. 
Parameters are taken from Ref.~\onlinecite{Scharf2019:PRB, 
Mayer2020:AEM}.}
\centering
\begin{tabular}{l|l|l}
Name						            & HgTe			& InSb 		    \\
\hline
\hline
Effective mass	($m^*$	)			    & 0.038$m_0$	& 0.013$m_0$ 	\\
Land\'{e} factor ($g^*$)			    & -10			& -20			\\
Induced SC gap	($\Delta_0$)			& 0.25 meV		& 0.21 meV		\\
SOC strength	($\lambda$)	            & 16 meV nm		& 15 meV nm		\\
Critical field at $0$K		            & 1.45 T		& 1.45 T		\\
Temperature	($T$)				        & 0.7 K			& 0.7 K			\\
Chemical potential in S ($\mu_S$)		& 1 meV			& 1 meV			\\
Chemical potential in N	($\mu_N$)	    & 1 meV			& 1 meV			\\
\hline
Junction width	($W_N$)			        & 96 nm		    & 96 nm		    \\
Left SC lead width	($W_S$)		        & 252 nm		& 252 nm		\\
Right SC lead width	($W_S$)		        & 252 nm  	    & 252 nm		\\
Junction length       ($L$)   	        & 4000 nm		& 4000 nm		\\
\hline
TB lattice constant	($a$)		        & 6 nm			& 6nm			\\
TB hopping parameter($t$)		        & 27.9 meV		& 81.5 meV		\\
\hline
\end{tabular}
\label{TBL:Parameters}
\end{table}


\begin{thebibliography}{112}
\bibitem [{not()}]{note1}%
  \BibitemOpen
  \href@noop {} {}\bibinfo {note} {In Ref.~\onlinecite{Pakizer2021:PRR} the
  magnetic field direction was defined with respect respect to the $[100]$
  crystallographic direction and denoted as $\theta_B$. Here we use the current
  direction as the axis with respect to which the magnetic field orientation,
  $\varphi_B$, is defined. The two definitions are connected through the
  relation $\varphi_B=\theta_B-\theta_c$, where $\theta_c$ indicates the
  direction of the current with respect to the $[100]$ direction.}\BibitemShut
  {Stop}%
\end{thebibliography}
\end{document}